\documentclass[twocolumn,showpacs,preprintnumbers,amsmath,amssymb]{revtex4}

\usepackage[dvips]{graphicx}
\usepackage{dcolumn}
\usepackage{bm}

\begin{document}


\title{Decoherence processes in a current biased dc SQUID}

\author{J. Claudon, A. Fay, L. P. L\'evy and O. Buisson}

\affiliation{CRTBT-LCMI, C.N.R.S.- Universit\'e
Joseph Fourier, BP 166, 38042 Grenoble-cedex 9, France}

\date{\today}

\begin{abstract}
A current bias dc SQUID behaves as an anharmonic quantum oscillator
controlled by a bias current and an applied magnetic flux. We consider
here its two level limit consisting of the two lower energy states $
\left| 0 \right>$ and $ \left| 1 \right> $. We have measured energy
relaxation times and microwave absorption for different bias currents
and fluxes in the low microwave power limit. Decoherence times are
extracted. The low frequency flux and current noise have been measured
independently by analyzing the probability of current switching from
the superconducting to the finite voltage state, as a function of
applied flux. The high frequency part of the current noise is derived
from the electromagnetic environment of the circuit. The decoherence 
of this quantum circuit can be fully accounted by these current and flux
noise sources.
\end{abstract}

\pacs{Valid PACS appear here}
\maketitle

In the past years, coherent manipulation of two and 
multi-level
quantum systems, efficient quantum readouts, entanglement between
quantum bits have been
achieved\cite{Yamamoto03,Vion02,Martinis02,Chiorescu03,Claudon04,Wallraff04}
demonstrating the full potential of quantum logic in solid state
physics. At present, future developments require longer
coherence times \cite{Ithier05,Bertet05,Astafiev04}.
In contrast with 
atomic system, the huge number of 
degree
of freedom makes its optimization a challenging
problem. Up to now, the most successful strategy has been to 
manipulate
the quantum system at particular working points where its coupling to
external noise is minimal\cite{Vion02}. Nevertheless, experimental 
analysis
of decoherence phenomena in superconducting circuits remains a 
priority for a full-control of
quantum experiments. Different models for
the noise sources have been proposed to describe the decoherence
processes acting on various
qubits\cite{Cooper04,Astafiev04,Bertet05,Ithier05}. However a 
complete and
consistent understanding of decoherence remains a current and open
problem. In this paper, we study decoherence processes of a 
phase qubit: the current biased dc SQUID.

\begin{figure}
\resizebox{0.45\textwidth}{!}{\includegraphics{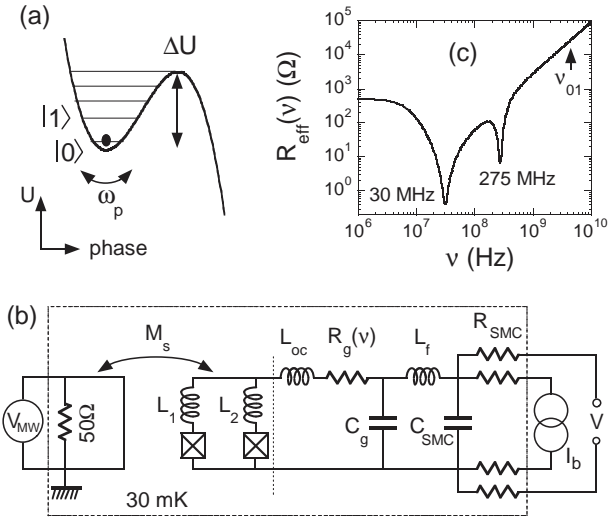}}
\caption{(a) Squid cubic-quadratic potential. (b) Electrical
environment of the SQUID.
(c) Calculated effective real impedance $R_{eff}$ versus frequency.}
\label{circuit}
\end{figure}

This superconducting circuit consists of two Josephson junctions 
(JJ), each with a
critical current $I_0$ and a capacitance $C_0$. The junctions are
embedded in a superconducting loop of inductance $L_s$, threaded by a
flux $\Phi_b$. In the limit where $L_s I_0 \approx \Phi_0 / 2\pi$, the
phase dynamics of the two junctions can be mapped onto a fictitious
particle following a one dimensional path in a 2D-potential
\cite{Claudon04}. If the biasing current $I_b$ is smaller than the
SQUID critical current $I_c$, the particle is trapped in a cubic
potential well characterized by its bottom frequency
$\omega_p(I_b,\Phi_b)$ and a barrier height $\Delta U(I_b,\Phi_b)$
(Fig.~\ref{circuit}.a). The quantum states in this anharmonic
potential are denoted $\left| n \right>$, with corresponding
energies $E_n$, $n=0,1,...$ In the following, only
the lowest states $\left| 0 \right>$ and $ \left| 1 \right>$ will be
involved. For $I_b$ well below $I_c$, these two levels are stable and
constitute a phase qubit.

When the bias current $I_b$ is close to $I_c$, $\Delta U$ decreases 
and
becomes of the order of a few $\hbar \omega_p$. The ground state can
tunnel through the potential barrier and the SQUID switches to a
voltage state\cite{Balestro03}. The tunnelling rate $\Gamma_0$ of the
ground state $\left| 0 \right>$ is given by the well-known MQT formula
for underdamped JJ\cite{Caldeira83}: $\Gamma_0(I_b,\Phi_b) = a 
\omega_p \exp(-36 \Delta
U/5 \hbar \omega_p)$, where $a$ is of order
unity.

The environment of the dc SQUID induces fluctuations of
the bias current and the bias flux. In this work, we show how the
current and flux noise sources can be separately quantified. This is 
achieved by escape measurements of the SQUID at specific working 
points where it is mostly sensitive to current or flux noise. Using 
these identified noise sources, the measured decoherence times are 
fit precisely as a  function of bias current.

Experimental results are analyzed assuming a linear coupling between
the SQUID and the environment degrees of freedom.  We suppose that
current $\widehat{\delta I}$ and flux $\widehat{\delta \Phi}$ noises
are generated by independent gaussian sources. Here, $\widehat{\delta
x} \ (x=I\ \text{or} \ \Phi)$ is an operator acting on the 
environment.
Their fluctuations are specified by the quantum spectral densities
$S_x(\nu)$\cite{dsp}.  In presence of flux microwave (MW)
excitation, the total Hamiltonian $\widehat{H}$ in the SQUID
eigenstates basis $\left\{ \left| 0 \right>, \left| 1 \right> 
\right\}$
reads: $\widehat{H} = - \frac{1}{2} h \nu_{01} \widehat{\sigma}_z  - h
\nu_R \cos(2 \pi \nu t) \widehat{\sigma}_x + \widehat{N}$ where
$\widehat{\sigma}_x$ and $\widehat{\sigma}_z$ are Pauli matrices and
$\nu_{01}=(E_1-E_0)/h$. The first term is the qubit Hamiltonian and 
the
second term describes the MW excitation of reduced amplitude $\nu_R$ 
at
frequency $\nu$. In this notation, $\nu_R$ is also the 
Rabi precession frequency
for a tuned excitation ($\nu = \nu_{01}$) . The last term is the 
coupling
to the noise sources. For our circuit it is, within linear response,
\begin{equation}
\begin{split}
\widehat{N} = - \frac{h}{2}  \widehat{\sigma}_x &\Big[
\frac{r_I(\theta)}{2\pi\sqrt{C_0 h \nu_{01}}} \widehat{\delta I} +
\frac{r_{\Phi}(\theta)}{\pi L_s\sqrt{C_0 h \nu_{01}}} \widehat{\delta
\Phi} \Big] \\ - \frac{h}{2} \widehat{\sigma}_z &\Big[
\Big(\frac{\partial \nu_{01}}{\partial I_b }\Big)  \widehat{\delta I}
+  \Big(\frac{\partial \nu_{01}}{ \partial \Phi_b}\Big)
\widehat{\delta \Phi} \Big].
\end{split}
\label{noise}
\end{equation}
where $\eta$ is the asymmetry inductance parameter (see below),
$r_I(\theta)=\cos\theta +\eta \sin\theta $,
$r_{\Phi}(\theta)=\sin\theta $ and $\theta$ is the angle between the
escape and the mean slope directions in the 2D
potential\cite{Lefevre92,Balestro03}. To first order, the transverse
noise proportional to $\widehat{\sigma}_x$ only induces 
depolarization. The
longitudinal term proportional to $\widehat{\sigma}_z$ induces "pure"
dephasing. The qubit sensitivity to longitudinal noise is
given by the partial derivatives $(\partial \nu_{01} /
\partial I_b)$ and $(\partial \nu_{01} / \partial \Phi_b)$.
They depend strongly on the experimental working point and increase
near the critical current.\\

The measured SQUID consists of two large aluminum JJs of $15 \:
\mu\text{m}^{2}$ area ($I_0 = 1.242 \: \mu\text{A}$ and $C_0 = 0.56 \:
\text{pF}$) enclosing a $350 \: \mu\text{m}^{2}$-area superconducting
loop. The two SQUID branches of inductances $L_{1}$ and $L_{2}$
contribute to the total loop inductance $L_s = 280 \: \text{pH}$ with
the asymmetry parameter $\eta=(L_{1}-L_{2})/L_{s}=0.414$. The 
immediate
electromagnetic environment of the SQUID is designed
to decouple the circuit from the external world. It 
consists of two cascaded LC filters (see
Fig.~\ref{circuit}.b). A large on-chip inductance $L_{oc}=9 \: 
\text{nH}$
is made of two long and thin superconducting wires which value, 
derived
from the normal state resistance, is dominated by the kinetic
inductance. The gold thin film parallel capacitor, $C_g \approx 150 \:
\text{pF}$, introduces a finite resistor. Its dc value  at $30 \:
\text{mK}$ is $R_g = 0.1 \: \Omega$ giving the gold resistivity
$\rho_{g}=1.2 10^{-8}\: \Omega$m.
The second
filter consists of the bounding wires, with an estimated inductance
$L_f = 3 \: \text{nH}$, and a surface mounted (SMC) capacitor $C_{\rm
SMC} = 2 \: \text{nF}$ and four $500 \: \Omega$ SMC resistors. The
nominal room temperature microwave signal is guided by $50 \: \Omega$
coaxial lines, attenuated at low temperature before reaching the SQUID
through a mutual inductance $M_s = 1.3 \: \text{pH}$. Special care was
taken in magnetic shielding and bias lines filtering. All 
these electrical parameters
were determined independently \cite{Balestro03,ClaudonThesis}.
Our environment model predicts two resonances (resistance dips in Fig.~\ref{circuit}.c).
They were observed in a similar set-up  and the associated resonance frequencies were in 
precise agreement with the model.

The current noise through the SQUID comes mostly from its immediate
environment thermalized at $T = 30 \: \text{mK}$ ($\nu_T \equiv k_{\rm
B}T/h = 600 \: \text{MHz}\ll \nu_{01}$). The quantum spectral density
of the current noise, $S_I(\nu)$ in this environment is set by the
fluctuation-dissipation theorem: $S_I(\nu) = h \nu \big[ \text{coth}
\big( \frac{h \nu}{2 k_B T} \big) + 1 \big] R_{\rm eff}(\nu)^{-1}$
where $R_{\rm eff}(\nu)^{-1}$ is the real part of the environment
circuit admittance. $R_{\rm eff}(\nu)$ is calculated using the
electrical circuit shown in Fig.~\ref{circuit}.b and is plotted in
Fig.~\ref{circuit}.c. To a good approximation, the root mean square
(RMS) current fluctuations are of order $\sqrt{k_B T/L_{\rm oc}}=6 \:
\text{nA}$. Most of the noise is peaked around $30 \: \text{MHz}$, a
frequency much smaller than $\nu_T$. A simple estimate of the flux
noise produced by the inductive coupling to the $50 \: \Omega$ coaxial
line shows it can be neglected in the following.

\begin{figure}
\resizebox{0.4\textwidth}{!}{\includegraphics{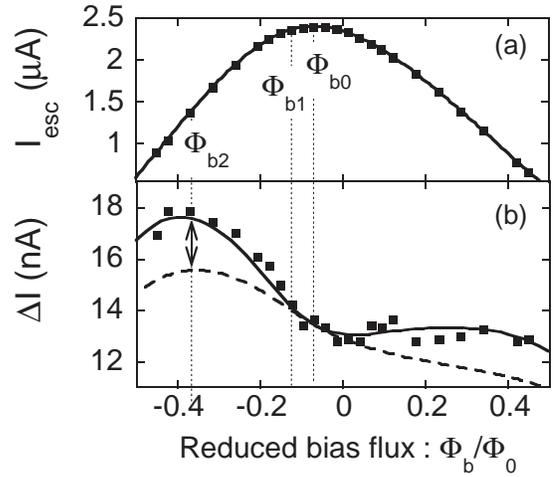}}
\caption{(a) Measured escape current (dots) versus external applied
flux fitted to MQT theory (solid line) at 30\:mK. (b) The width of the
probability distribution $P_{esc}(I_{b})$ (dots) fitted to the 2D MQT
predictions. The solid curve takes the low frequency flux noise into
account while the dashed line does not. At bias flux $\Phi_{b0}$ 
(resp.
$\Phi_{b1}$) the sensitivity to flux noise is zero (resp. small) while
it is maximum at $\Phi_{b2}$.} \label{MQT}
\end{figure}

The escape probability $P_{\rm esc}(I_{b})$ out of the superconducting
states is measured at fixed flux using dc current pulses with 
$\Delta t = 50 \:
\mu\text{s}$ duration and $I_{b}$ amplitude. Each measurement 
involves 5000 identical current pulses and the total acquisition time 
is $T_{\rm m} = 10\:$s. The escape current $I_{\rm esc}$ is
defined as the current $I_b$ where the escape probability
$P_{esc}(I_b)=0.5$ and the width of the switching curve $\Delta
I=I_h-I_l$ as the difference between the currents where
$P_{esc}(I_h)=0.9$ and $P_{esc}(I_l)=0.1$. In Fig.~\ref{MQT}, the
dependence of $I_{\rm esc}$ and $\Delta I$ on $\Phi_b$ are plotted. By
fitting the escape current curve $I_{\rm esc}(\Phi_b)$, the
experimental parameters of the SQUID $(I_0,C_0,L_s,\eta)$ are
determined.

Moreover, escape measurements are a sensitive tool to 
characterize noise (frequency range and amplitude). If noise 
frequencies exceed the
inverse of a current pulse duration $\Delta t^{-1}$, the tunnel rate
fluctuates during each current pulse. The escape probability is
controlled by the average $\big< \Gamma_0 \big>$ escape rate in the
frequency window [$\Delta t^{-1}$, $\nu_T$]: $P_{\rm esc} = 1 - \exp
\big[ - \big< \Gamma_0 \big( I_b+\delta I,\Phi_b+\delta \Phi \big)
\big> \Delta t \big]$ \cite{Martinis88,Balestro03}.  The current noise
produced by the electrical environment lies in this frequency 
interval. Its effect is to
decrease $I_{\rm esc}(\Phi_b)$ by about $ 6 \: \text{nA}$, the RMS
current fluctuations (unobservable in Fig.~\ref{MQT}.a). Similarly, 
the
width of the switching curve is not affected.

On the other hand, if noise frequencies are slower than $\Delta
t^{-1}$, the tunnel rate is constant during a pulse, but fluctuates
from pulse to pulse. In this limit, the escape probability becomes
$P_{\rm esc}= \big<1 - \exp \big[ - \Gamma_0 \big( I_b + \delta
I,\Phi_b + \delta \Phi \big) \Delta t \big] \big>$, where the
statistical average $\big< \big>$ is now in frequency range from
$T_{\rm m}^{-1}$ to $\Delta t^{-1}$. To first order, low frequency
noise does not affect $I_{\rm esc}$, but increases the width $\Delta
I$. Thus $\Delta I$ is the best quantity to probe the origin and the
magnitude of the low frequency fluctuations: if the flux $\Phi_b$ is
set at the value $\Phi_{b0}$ which maximizes $I_c$, the SQUID is only
sensitive to current fluctuations since $\frac{\partial
\nu_{01}}{\partial \Phi_b}=0$. In the vicinity of this flux, the
measured width is explained by the usual MQT theory.   Hence the
measured RMS current fluctuations in the [$T_{\rm m}^{-1}$, $\Delta
t^{-1}$] interval (low frequency current noise) is below $0.5 \:
\text{nA}$, the error bar in $\Delta I$ measurements. This is
consistent with the $0.1 \: \text{nA}$ RMS value derived from the
spectral density of noise at frequencies below $\Delta t^{-1}$. For
other applied fluxes, the width is slightly larger than MQT 
prediction,
indicating a residual low frequency flux noise. The dependence of
$\Delta I$ on $\Phi_b$ shown in Fig.~\ref{MQT}.b is perfectly 
explained
by a gaussian low frequency flux noise. Its RMS amplitude, $\big<
\delta \Phi_{LF}^2 \big>^{1/2} = 5.5 \times 10^{-4} \: \Phi_0$, is
extracted from the fit shown in Fig.~\ref{MQT}.b and is attributed to 
the
flux noise in the $[100 \: \text{mHz},20 \: \text{kHz}]$ frequency
interval. The origin of flux noise may be due
to vortices
trapped in the four aluminum
contact pads located at a $0.5 \: \text{mm}$ distance from the 
SQUID.

\begin{figure}
\resizebox{0.4\textwidth}{!}{\includegraphics{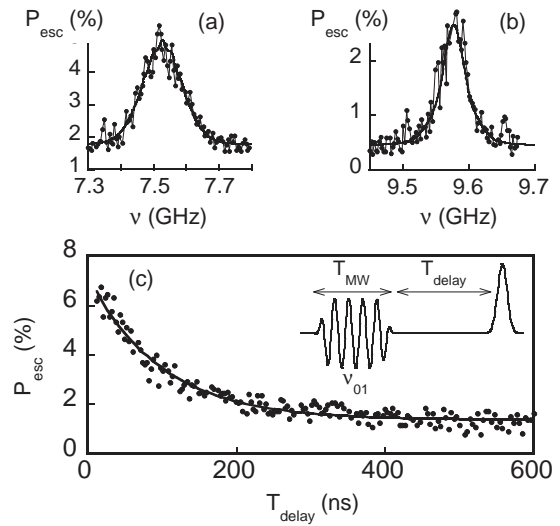}}
\caption{(a) and (b) Escape probability versus applied microwave
frequency with amplitude $\nu_{R} < 5 \ \text{MHz}$ at two different
working points $(I_b=2.288 \: \mu\text{A},\Phi_{b1} = -0.117 \:
\Phi_0)$ and $(I_b=0.946 \: \mu\text{A},\Phi_{b2} = - 0.368 \:
\Phi_0)$, respectively. The points are experimental data and the
continuous lines are the Fourier transforms of $f_{\rm coh}(t)$ (see
text). (c) Measured escape probability versus delay time (dots) fitted
to an exponential law with $T_{1}= 95 \: \text{ns}$ (continuous line).
The inset specifies the timing of the measurement pulse which follows 
the
MW excitation pulse.} \label{spectroscopy}
\end{figure}

\begin{figure}
\resizebox{0.4\textwidth}{!}{\includegraphics{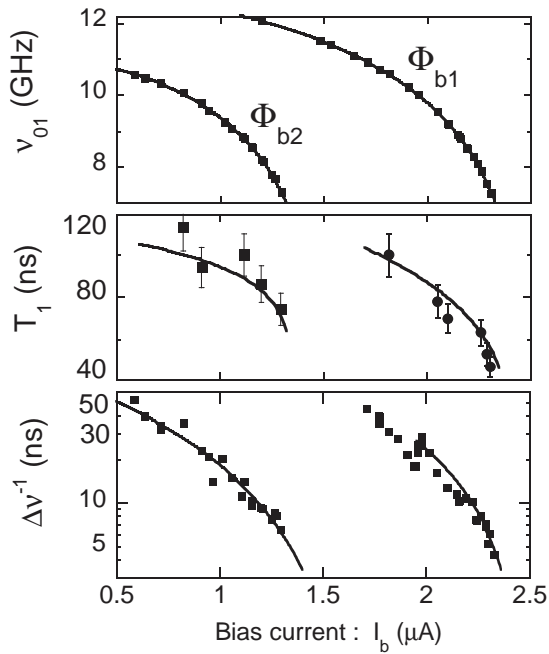}}
\caption{Resonant transition frequency (a), relaxation time (b) and
microwave width (c) as function of bias current at $\Phi_{b1}
= -0.117 \: \Phi_0 $  and $\Phi_{b2} = - 0.368 \: \Phi_0$ respectively
right and left curves. Symbols correspond
to experiments and continuous line to model predictions. }
\label{decoherence}
\end{figure}

Hereafter we discuss dephasing and relaxation induced by the
noise sources previously identified. These incoherent processes are 
experimentaly studied with low power spectroscopy and energy 
relaxation measurement. As described in
Ref.\cite{Claudon04}, a MW flux pulse is applied followed by a $2 \: 
\text{ns}$
 duration dc flux pulse to perform a fast but adiabatic
measurement of the quantum state of the SQUID 
(Fig.~\ref{spectroscopy}.c
inset). The duration $T_{MW}=300 \: \text{ns}$ of MW pulses is
sufficient to reach the stationnary state where the population $p_1$ 
of
the level $\left|1\right>$ only depends on $\nu$ and the amplitude 
$\nu_R$. The microwave amplitude $\nu_R$ is calibrated using Rabi 
like oscillations\cite{Claudon04}. In the two level experiments 
discussed in this paper, 
the measured escape probability $P_{\rm esc}$ induced by the dc flux 
pulse can be interpreted as  $P_{\rm esc}= P_{\rm
esc}^{\left|0\right>}+(P_{\rm esc}^{\left|1\right>}-P_{\rm
esc}^{\left|0\right>}) \times p_1(\nu,\nu_R)$. $P_{\rm 
esc}^{\left|n\right>}$ denotes the escape probability out of the
pure state $\left|n\right>$. 
In Fig.~\ref{spectroscopy}.a and \ref{spectroscopy}.b, the escape 
probability versus
microwave frequency $\nu$ are plotted at two different biasing 
points. The
experimental curves present a resonant peak which position and full 
width at half maximum 
define the resonant frequency $\nu_{01}$ and $\Delta \nu$. 
Spectroscopy experiments are performed in the linear regime and 
$\Delta\nu$ is experimentally checked to be independent of the MW
amplitude. Relaxation measurements were performed by populating the 
$\left| 1 \right>$ state
with low power MW tuned at $\nu_{01}$ during a time $T_{\rm MW}=300 \:
ns$, and measuring its population with increasing time delay $T_{\rm
delay}$ after the end of the MW pulse. As shown in
Fig.~\ref{spectroscopy}.c, the escape probability follows an
exponential relaxation with a characteristic time $T_1$. In 
Fig.~\ref{decoherence}, measured resonant frequency $\nu_{01}$,
relaxation time $T_1$ and the inverse of microwave width
$\Delta\nu^{-1}$ are plotted versus current bias for the two different
applied fluxes $\Phi_{b1}$ (close to $\Phi_{b0}$) and $\Phi_{b2}$ 
shown
in Fig.~\ref{MQT}. $\nu_{01}$, $T_1$ and $\Delta\nu^{-1}$ decreases as
$I_{b}$ gets closer to $I_{c}$. For these two applied fluxes, the
$\nu_{01}$ dependence fits perfectly the semiclassical formulas for a
cubic potential \cite{Larkin86} using the same SQUID electrical
parameters as those extracted from escape measurements.

The depolarization rate $T_1^{-1}$ is given by the sum
$T_1^{-1}=\Gamma_R+\Gamma_E$ of the relaxation $\Gamma_R$ and the
excitation $\Gamma_E$ rates. These two rates are calculated using 
Fermi
golden rule. At low temperature, excitation can be neglected and
$\Gamma_R$ reads:
$$ \Gamma_R = \frac{r_I^2(\theta)}{4C_0 h \nu_{01}} S_I(\nu_{01}) +
\frac{ r_\Phi^2(\theta)}{L_s^2 C_0 h \nu_{01}} S_{\Phi}(\nu_{01}).$$
Neglecting the high frequency part of the flux noise, one obtains $T_1
= 2 R_{\rm eff}(\nu_{01}) C_0/r_I^2(\theta)$ where $R_{\rm eff}(\nu_{01}) 
=(2 \pi L_{oc} \nu_{01})^2 / R_g(\nu_{01})$.  $R_g(\nu)= \alpha
R_{s}$ is the high frequency resistance of the gold capacitor
where  $R_{s}=\sqrt{\pi \mu_{0}\rho_{g} \nu}$ is the surface resistance and  $\alpha$ is
a dimensionless
geometrical parameter. The $T_1$ versus $I_{b}$ dependence is well fitted with $\alpha=200$
as the only adjustable parameter (Fig.~\ref{decoherence}.b). 
This value is the right order of magnitude for the geometry of the gold capacitor.

Relaxation alone is too weak to explain the value of $\Delta\nu$ and
"pure" dephasing also contribute to the linewidths.
 First, we consider the time evolution of the reduced density matrix
in the basis $\left\{ \left| 0 \right>, \left| 1 \right> \right\}$
in the absence of MW. The linear coupling to noise sources induces a
time decay of the amplitude $f_{\rm coh}(t)$ of the coherence terms.
Since current and flux noises are independent, $f_{\rm coh}(t)$ is
factorized as $f_{\rm coh}(t)=f_I(t)\, f_{\Phi}(t) \exp (-2t/T_1)$
where $f_I(t)$ and $f_{\Phi}(t)$ are respectively the "pure" dephasing
contributions due to current and flux noises.

The current contribution is given by the well-known gaussian noise
formula \cite{Cottet02,Ithier05}: $f_{I}(t) = \exp \big[
-\frac{1}{2} t^2 \times \big( 2 \pi \frac{\partial
\nu_{01}}{\partial I_b } \big)^2 \int_{-\infty}^{+\infty} d\nu
S_I(\nu) \text{sinc} (\pi \nu t) \big]$, where $(\partial \nu_{01} /
\partial I_b)$ is extracted directly from the slope of the 
experimental curve of
Fig.~\ref{decoherence}.a. We neglect flux noise contributions with
frequencies higher than $20 \: \text{kHz}$.
 Since the acquisition time of absorption spectra and escape
measurements are similar, the SQUID undergoes the same RMS flux
fluctuations in the two experiments. In these conditions,
$f_{\Phi}(t)$ takes the simple gaussian form: $f_{\Phi}(t) = \exp
\big[ -\frac{1}{2} t^2  \times \big(2\pi \frac{\partial
\nu_{01}}{\partial \Phi_b } \big)^2 \big< \delta \Phi_{LF}^2 \big>
\big]$. $(\partial \nu_{01} / \partial \Phi_b)$ was computed using
the known electrical parameters of the SQUID\cite{hysteresys}.

Within linear response, the shape of the resonance curve
is proportional to the Fourier transform (FT) of $f_{\rm coh}(t)$.
Resonance curves in Fig.~\ref{spectroscopy}.a and 
\ref{spectroscopy}.b are fitted using
$P_{esc} - P_{esc}^{\left|0\right>} \propto 
\text{FT}\{f_{coh}\}(\nu-\nu_{01})$ (continuous line).
Our model explains perfectly the shape of the experimental curves. In
Fig.~\ref{decoherence}.c, the theoretical width $\Delta \nu$ extracted
from the curve $\text{FT}\{f_{coh}\}(\nu)$ is in very good agreement
with experimental data without free parameter. When $I_b$ gets close 
to $I_c$, the partial derivatives
$(\partial \nu_{01} /
\partial I_b)$ and $(\partial \nu_{01} /
\partial \Phi_b)$ increase: the noise sensitivity increases
and $\Delta \nu$ broadens. For bias points corresponding to
$\Phi_{b2}$, the width is due to current and flux noise.  For a bias
flux equal to $\Phi_{b1}$, the effect of flux noise is small and the
width is dominated by current noise.  At this flux, for $I_b < 1.95\:
\mu\text{A}$, our model predicts satellite resonances around 
$\nu_{01}$
which are not observed. Other noise mechanism may blur the predicted
satellite peaks.

In conclusion, we have shown how the flux and current noise present in
this controlled quantum circuit can be separately identified. We
measured the decoherence times at low microwave power where the 
quantum
circuit can be reduced to a two level system. Analyzing the coupling 
of
the SQUID to the known noise sources, the measured relaxation times 
and
the resonance width can be fully understood.

We thank E. Colin, V. H. Dao, K. Hasselbach, F.W.J Hekking, B. 
Pannetier, P. E. Roche,
J. Schrieff, A. Shnirman for very useful discussions. This work was
supported by two ACI programs and by the Institut de Physique de la
Mati\`ere Condens\'ee.


\begin{thebibliography}{}
\bibitem{dsp} The quantum spectral density is defined as follows :
$S_x(\nu)=\int dt \big< \widehat{\delta x}(0)\widehat{\delta x}(\tau)
\big> e^{-i 2 \pi \nu \tau}$.
\bibitem{hysteresys} $(\partial \nu_{01} /\partial \Phi_b)$ could be 
measured by changing slighlty $\Phi_b$. In fact, 
We kept $\Phi_b$ as constant as possible during experiments, because 
of a static flux hysteresis about $10^{-3} \: \Phi_0$.
\end{thebibliography}
\end{document}